# Hydrogen Compounds of Group-IV Nanosheets

L. C. Lew Yan Voon* and E. Sandberg

*Department of Physics, Wright State University,*

*3640 Colonel Glenn Hwy, Dayton, OH 45435*

R. S. Aga

*Department of Chemistry, Wright State University,*

*3640 Colonel Glenn Hwy, Dayton, OH 45435*

A. A. Farajian

*Department of Mechanical and Materials Engineering,*

*Wright State University, 3640 Colonel Glenn Hwy, Dayton, OH 45435*

(Dated: July 2, 2010)

## Abstract

The structural and electronic properties of the hydrides of silicene and germanene have been studied using ab initio calculations. The trend for the M-H (M=C, Si, Ge) bond lengths, and corresponding bond energies, is consistent with the atomic size trend, and comparable to those of $MH_4$ hydrides. Band structures were also obtained for the buckled configuration, which is the stable form for both silicene and germanene. Upon hydrogenation, both silicane (indirect gap) and germanane (direct gap) are semiconducting.

Since 2004, graphene has caused a revolution in materials research due primarily to its two-dimensional nature and to the linear dispersion of its band structure near the Dirac point[1]. These features convey to graphene some remarkable properties such as displaying Klein tunneling, the anomalous integer quantum Hall effect for relatively low magnetic fields at room temperature and a finite minimum DC conductivity[2]. More recently, hydrogenation has been achieved[3] leading not only to a new material, graphane[4], but also to prospects for application such as for hydrogen storage. Investigating similar characteristics of other group-IV elements such as silicon and germanium is, therefore, important and promising since silane and germane exist alongside methane. It turns out that the study of the stability of a single hexagonal sheet of Si and Ge goes back at least to a 1994 paper by Takeda and Shiraishi[5] using density-functional theory (DFT) within the local-density approximation (LDA). The fundamental result is that the hexagonal sheet of Si and Ge would rather be buckled than flat, reflecting the reduced tendency of the group-IV elements other than carbon to form $sp^2$ bonding[6]. In analogy to the study of graphene, there have been a few other papers on the single sheet of silicon towards making silicon nanotubes[7,8] and the single sheet of silicon has been named silicene[9]. More importantly, even the buckled sheet has been shown to possess a zero gap and Dirac points[9]. In these early work on silicene, both Durgun *et al.* and Yang and Ni used DFT and the generalized gradient approximation (GGA) while Guzmán-Verri and Lew Yan Voon used the empirical tight-binding method; Durgun *et al.* found the buckling height to be 0.45 Å. Another structure considered theoretically is that of a flat and a corrugated bilayer of Si sandwiched between monolayers of molybdenum[10], $(MoSi_{12})_n$, and is predicted to have a generalized gradient approximation (GGA) indirect gap of $\sim 0.5$ eV. More recently, interest has grown in the study of such sheets including Group IV-IV and III-V compounds[11–19]. Just as for graphene, the culmination is, without doubt, the experimental growth of silicene nanosheets[20] and nanoribbons[21–23]. Thus, already in 2006, Nakano *et al.* reported the synthesis of single-crystal silicon monolayer sheets via the chemical exfoliation of $CaSi_2$. In the more recent work, the Si nanoribbons were deposited on a silver substrate[21,23]. For Ge, Cahangirov *et al.*[12] found two local minima with buckling heights of 0.64 Å (low buckled) and $\approx 2.0$ Å (high buckled) but with only the former being stable. There are other papers on the two-dimensional Ge sheet named germanene[17]; however, we note that three of them studied the flat germanene structure as if it is the stable one[11,17,19]; hence, their usefulness is limited. Beyond the lattice properties, the electronic and phonon structures have also been studied. Thus, graphene, silicene and germanene are all predicted to be zero-gap semiconductors.

In this letter, we present the properties of silicene and germanene analogs of graphane. It is already known that all group-IV elements form $MH_4$ hydrides[6]. It is also known that $sp^2$ (and $sp$) bonding is only strongly favored for hydrocarbons. We note that, while graphene consists of a flat hexagonal sheet, graphane has a buckled structure as its most stable form. Hence, it is interesting to investigate the structure and electronic properties of the hydrogenated silicene and germanene sheets. For convenience, we will label the flat sheets $\alpha$-type while the buckled sheets will be labeled $\beta$-type.

The sheets are modeled using a hexagonal unit cell in the plane with two atoms and a superperiodicity of at least 20 Å in the perpendicular direction forming a supercell. All of our calculations were completed using the ABINIT DFT package[24,25] with pseudo-potentials and a plane wave basis set. We completed calculations using both LDA and GGA Troullier-Martin pseudopotentials. Convergence tests showed that a 40 Hartree cutoff was sufficient for all of our calculations. We optimized the geometry of the systems using the Broyden-Fletcher-Goldfarb-Shanno minimization scheme, and a $10^{-4}$ Hartree/Bohr tolerance on the maximal force on the atoms. We allowed smearing of the Energy cutoff by 0.5 Hartree, and a maximal scaling of the lattice dimensions of 1.1. The supercell's shape, volume, and ionic positions were all given the freedom to relax. We used a $10^{-8}$ Hartree tolerance on the difference of total energy which when reached twice in succession ended an SCF cycle. We used a macroscopic dielectric constant of 12 (which is standard for semiconductors) to speed the convergence of the SCF cycles. The k-point grid was generated using a 2×2×2 Monkhorst-Pack grid, based on the primitive vectors of the cell, which ABINIT then shifted four times. Increasing the density of the grid proved not to increase the convergence of the total energy.

Most of the relevant structural parameters computed are given in Table I, where $a$ is the in-plane hexagonal lattice constant and $\Delta z$ is the buckling height. It was found that all the structures were buckled except for pure graphene (Table I). Without hydrogen this reflects the already known fact that, except for carbon, the group-IV elements do not like to form $sp^2$ bonding. For Si, the buckling height is found to be 0.44 Å without hydrogen and 0.72 Å with hydrogen. The hydrogen alternates above and below the plane; this is similar to graphane. This is, of course, the configuration one would get from $sp^3$ bonding. However, without hydrogen, pure $sp^3$ bonding would lead to alternate Si atoms being 0.78 Å below the plane. Hence, the height obtained for silicene is less indicating a behaviour in between $sp^2$ and $sp^3$. Our value of 0.44 Å agrees with most other published results[8,12] but differs slightly from that reported by Ding and Ni using SIESTA, an atomic orbital based method. While the difference is larger than typical lattice parameter differences among various *ab initio* calculations, direct comparison is difficult since the methodology is different. Similar to Cahangirov and coworkers[12], we found that, for Ge, there are two local minima [which they labeled low buckled (LB) and high buckled (HB)]; through phonon calculations, they argued that the HB is unstable. The M-H bond length is very similar to the corresponding bonding for the MH$_4$ hydride and the increase down the periodic table reflects the increasing atomic size.

TABLE I: Structural parameters. The $\alpha$ ($\beta$) structure refers to the flat (buckled) sheet and LB (HB) stands for low (high) buckled.

|  | $a$ (Å) | $\Delta z$ (Å) | M-H (Å) | MH$_4$[6] |
|---|---|---|---|---|
| $\alpha$-C | 2.4496 | 0 | | |
| $\beta$-CH | 2.5137 | 0.45 | 1.0835 | 1.0870 |
| $\beta$-Si | 3.8202 | 0.44 | | |
| | | 0.44[12], 0.53[13] | | |
| $\beta$-SiH | 3.8202 | 0.72 | 1.5016 | 1.48 |
| $\beta$-Ge (LB) | 4.0 | 0.71 | | |
| | | 0.64[12] | | |
| $\beta$-GeH (LB) | | 0.69 | 1.53 | 1.5251 |
| $\beta$-Ge (HB) | 2.9042 | 1.84 | | |
| | | $\approx 2$[12] | | |
| $\beta$-GeH (HB) | | 1.84 | 1.6482 | |

An understanding of the structure can be obtained by studying the various energy contributions to the total energy of $\beta$-Ge sheets compared to $\alpha$-Ge are given in Fig. 1. The $\alpha$-Ge structure is the local minimum structure obtained by imposing a flat sheet constraint. The individual contributions are similar to the results obtained by Takeda and Shiraishi[5] for Si and Ge. This shows that the larger atomic separation of the buckled state is responsible for the energy stabilization with respect to the flat sheet, mainly by reducing the repulsive lattice and Hartree terms.

Bond energies were calculated as follows:

$$\frac{1}{2}\left[E_{\text{total}}(MH) - E_{\text{total}}(M) - 2E_{\text{total}}(H)\right], \tag{1}$$

where, for example, $E_{\text{total}}(H)$ is the total energy of a lone hydrogen atom (computed in a 30 Bohr supercell and the calculation was spin polarized) and is, here, found to be $-12.99\,\text{eV}$. The various bond energies are given in Table II and compared to the bond energies of the well-known hydrides[6]. It can be seen that the trends are identical. The magnitudes reflect the covalent nature of the bonds and the decrease in bond energy down the table can be explained in terms of the larger cationic radii. Since, so far, such two-dimensional hydrocompounds have not been identified (as opposed to the three-dimensional hydrides), one might suggest a more likely synthesis route via the initial formation of the pure two-dimensional sheets before hydrogenation.

TABLE II: Bond energies.

|  | $E$(MH) (eV) | $E$(M) (eV) | Bond (eV) | $MH_4$[6] |
|---|---|---|---|---|
| C-H | -360.007 | -326.672 | -3.678 | -4.306 |
| Si-H | -272.172 | -239.873 | -3.160 | -3.317 |
| Ge-H (LB) | -305.68 | -274.109 | -2.8 | -2.992 |

The band structures are given in Figs. 2–4. The graphene ones are provided to show consistency with already published results. As proven for silicene sheets[9], the sheets retain the Dirac point even when hydrogenated. The reason is that neither the buckling[9] nor the hydrogenation reduces the symmetry responsible for the zero gap. Our results for the pure sheets agree with recently published band structures[12]. We note that the germanene band structures of Lebègue and Eriksson[11], of Houssa et al.[17] and of Suzuki and Yokomizo[19] correspond to the flat sheet, which is not the stable minimum structure. Hydrogenation opens a gap for both silicene and germanene (as it does for graphene). For "silicane", the silicene sheet with complete hydrogenation, the LDA gap obtained is about $2\,\text{eV}$; hence, the actual gap might be larger. Note, though, that the minimum gap is indirect. Similarly, the LDA gap for "germanane", the hydrogenated germanene sheet, is about $1.5\,\text{eV}$ and the gap is direct. Compared to LDA calculations for graphane, the latter has a band gap of about 3.5 eV. Thus, it is expected that the correct band gap for silicane should be less than that of graphane and in the 3-4 eV range, and for germanane to be closer to 3 eV.

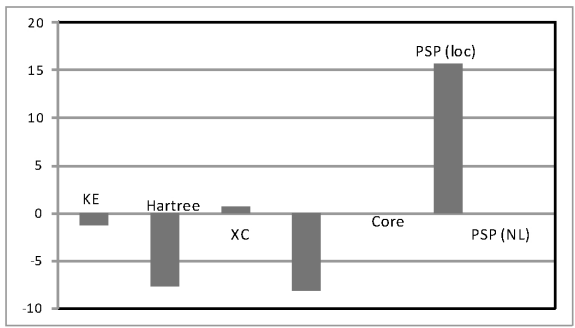

FIG. 1: Various energy contributions (in Hartree) to the total energy of $\beta$-Ge (LB) sheets compared to $\alpha$-Ge. XC denotes exchange-correlation, KE kinetic energy, and PSP pseudopotential.

In summary, we used density-functional theory to calculate structures, total energies and band structures of group-IV (C, Si, Ge) nanosheets with and without hydrogen. The structures were all buckled except for pure graphene. M-H bond energies weaken down the periodic table but are otherwise comparable to the corresponding hydride ones. It is, therefore, expected that the two-dimensional nanosheet hydrides would form stable compounds that could be synthesized; this has already been achieved for graphane. Both silicane (SiH) and germanane (GeH) are predicted to be semiconducting with relatively large gaps near 3 eV, which is about half of that of graphane. The Dirac point is preserved even upon hydrogenation.

The present results have been obtained through the use of the ABINIT code, a common project of the Université Catholique de Louvain, Corning Incorporated, and other contributors (URL http://www.abinit.org). This work was supported in part by an allocation of computing time from the Ohio Supercomputer Center. The authors would also like to thank the Ohio Board of Regents for partial support through a Research Challenge grant.

---

* Corresponding author: lok.lewyanvoon@wright.edu

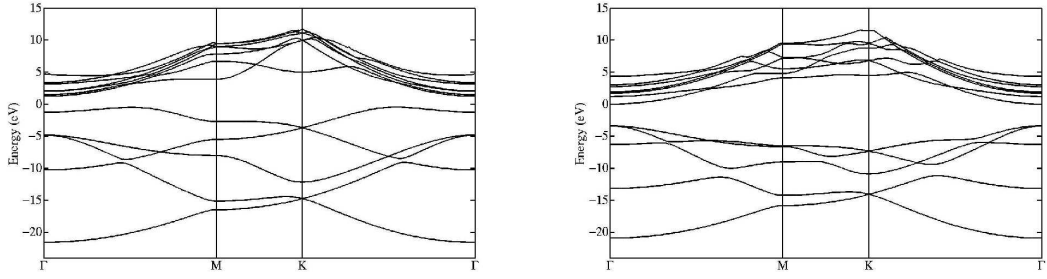

FIG. 2: Band structure of graphene sheet (a) without and (b) with hydrogen.

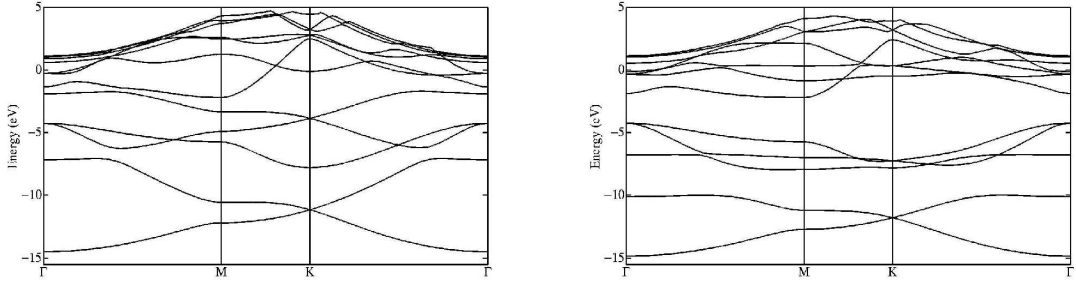

FIG. 3: Band structure of buckled silicene sheet (a) without and (b) with hydrogen.

orieva, and A. A. Firsov, Science, **306**, 666 (2004).

[2] A. H. C. Neto, F. Guinea, N. M. R. Peres, K. S. Novoselov, and A. K. Geim, Reviews of Modern Physics, **81**, 109 (2009).

[3] D. C. Elias, R. R. Nair, T. M. G. Mohiuddin, S. V. Morozov, P. Blake, M. P. Halsall, A. C. Ferrari, D. W. Boukhvalov, M. I. Katsnelson, A. K. Geim, and K. S. Novoselov, Science, **323**,

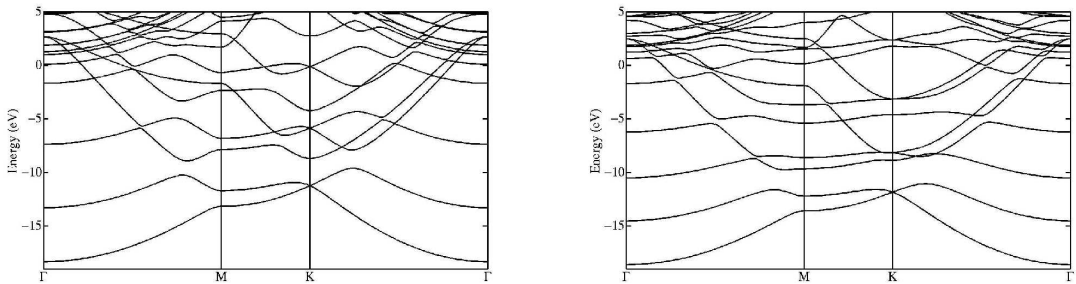

FIG. 4: Band structure of high-buckled germanene sheet (a) without and (b) with hydrogen.